\documentclass[reprint,aip,jcp,amsmath]{revtex4-1}

\usepackage{graphicx}

\newcommand{\Ket}[1]{\vert#1\rangle}
\newcommand{\Bra}[1]{\langle#1\vert}
\newcommand{\T}[1]{\mathrm{#1}}

\newcommand{\Op}[1]{\hat{#1}}
\newcommand{\BS}[1]{({#1})}
\newcommand{\diff}{\mathrm{d}}

\begin{document}

\title{Vibronic exciton theory of singlet fission.~III.~How vibronic coupling and thermodynamics promote rapid triplet generation in pentacene crystals}

\author{Roel Tempelaar}
\email{r.tempelaar@gmail.com}
\author{David R.~Reichman}
\email{drr2103@columbia.edu}
\affiliation{Department of Chemistry, Columbia University, 3000 Broadway, New York, New York 10027, USA}

\begin{abstract}
We extend the vibronic exciton theory introduced in our previous work to study singlet fission dynamics, in particular addressing recent indications of the importance of vibronic coupling in this process. A microscopic and non-perturbative treatment of electronic and selected vibrational degrees of freedom in combination with Redfield theory allows us to dynamically consider clusters of molecules under conditions close to those in molecular crystals that exhibit fission. Using bulk pentacene as a concrete example, our results identify a number of factors that render fission rapid and effective. Strong coupling to high-frequency Holstein modes generates resonances between the photo-prepared singlet and product triplet states. We furthermore find the large number of triplet combinations associated with bulk periodic systems to be critical to the fission process under such vibronically resonant conditions. In addition, we present results including, in an approximate manner, the effects of Peierls coupling, indicating that this factor can both enhance and suppress fission depending on its interplay with vibronic resonance and thermodynamics.
\end{abstract}

\maketitle

\section{Introduction}

Singlet fission,\cite{Smith_10a, Smith_13a} the conversion of a singlet excited state into two triplet excitons in molecular materials, both challenges our fundamental understanding of the photophysics of solids and molecules, and holds the technological promise to circumvent\cite{Hanna_06a} the Shockley-Queisser efficiency limit for simple solar cells.\cite{Shockley_61a} Near energetic resonance between the singlet and triplet pair states as well as spin entanglement of triplet pair intermediates have been recognized as crucial ingredients in singlet fission, but a complete set of design rules for fission materials remains lacking despite recent progress.\cite{Smith_10a, Smith_13a, Rao_17a} The limitations of our understanding of singlet fission are perhaps best exemplified by the surprisingly short singlet-to-triplet conversion time constant in crystalline pentacene, which through time-resolved spectroscopic measurements has been observed to lie within 150 fs,\cite{Chan_11a} or $\leq100$ fs,\cite{Wilson_11a, Bakulin_16a} for which a microscopic explanation remains to be found. Arguably, the closest agreement so far based on microscopic modeling of extended crystals was provided by Berkelbach \textit{et al.},\cite{Berkelbach_14a} yielding a time constant of 270~fs. However, the lack of quantitative agreement with experiment offers the opportunity to identify mechanistic principles, potentially omitted in the original modeling that impact singlet fission and whose importance extends beyond the description of bulk pentacene.

{Recent time-resolved spectroscopic experiments\cite{Musser_15a, Bakulin_16a, Monahan_16a} provided indications that coupling of vibrational modes to electronic transition energies plays a key role in facilitating rapid fission, an idea that can be traced back to the 1970s.\cite{Swenberg_74a}} Such Holstein-type vibronic coupling was treated perturbatively by Berkelbach \textit{et al.},\cite{Berkelbach_14a} but subsequent work has shown that the non-perturbative nature of the coupling to select modes enhances the mixing between the singlet and triplet pair states.\cite{Bakulin_16a, Fujihashi_17a, Morrison_17a} Another factor that has recently resurfaced in the discussion of singlet fission is Peierls-type vibronic coupling,\cite{Castellanos_17a} whereby the vibrational modulation of the electronic interactions between molecules occurs. Both Holstein and Peierls coupling have been theoretically demonstrated to benefit the intramolecular fission rate in chemically linked pentacenes.\cite{Huang_17a} Nevertheless, non-perturbative treatments of these mechanisms in a crystalline setting have remained limited to phenomenological models based on few-state systems,\cite{Bakulin_16a, Monahan_16a, Fujihashi_17a, Morrison_17a, Castellanos_17a} in which the potentially extended character of the involved excited states are not accounted for. As such, the significance of such modes for capturing the realistic, microscopic behavior representative of extended crystals remains to be verified.

\begin{figure}
\includegraphics{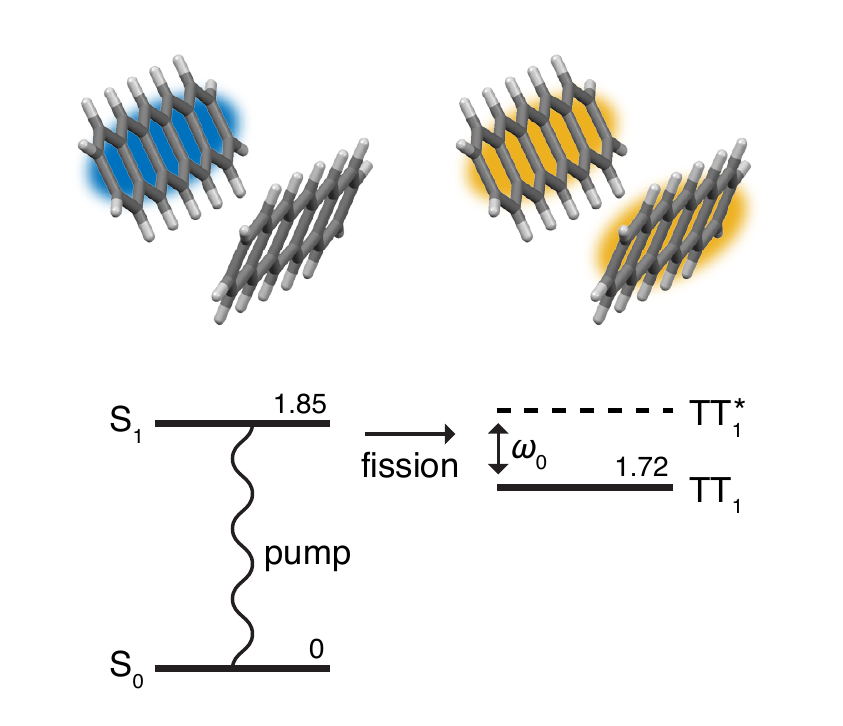}
\caption{Energy level diagram of the adiabatic excited states involved in singlet fission in crystalline pentacene, as suggested by experiment\cite{Bakulin_16a} and confirmed through microscopic modeling in the preceding articles of this series.\cite{Tempelaar_17a, Tempelaar_17b} Optical excitation (pump) couples between the (singlet) ground state, S$_0$, and the singlet excited state, S$_1$. From there, fission populates the triplet pair state, TT$_1$, possibly through its variant dressed by a single vibrational quantum of mode $\omega_0$, labeled TT$_1^*$. Numbers indicate energies (in eV) previously obtained for pentacene crystals.\cite{Tempelaar_17a} Shown on top is a schematic of the diabatic singlet and triplet pair excitations based on a pair of pentacene molecules.}
\label{Fig_Scheme}
\end{figure}

In the first paper of this series, hereafter referred to as Paper I,\cite{Tempelaar_17a} we sought to bridge this critical gap in the theoretical literature by formulating a microscopic model in which electronic degrees of freedom interacting with selected vibrational modes are treated non-perturbatively and at the molecular level. The model was parametrized for pentacene single crystals based on information from first principles calculations and experiments,\cite{Holmes_99a, Tsiper_03a, Yamagata_11a, Beljonne_13a, Berkelbach_13b, Hestand_15a} and validated by comparing simulations of the polarization-dependent linear absorption spectrum to experimental measurements.\cite{Hestand_15a} For each molecule, a single vibration was included involving a quantum of 1380~cm$^{-1}$, representative of a high-frequency, symmetric stretching mode of frequency $\omega_0$. The picture that emerged from this model is schematically illustrated in Fig.~\ref{Fig_Scheme}. Consistent with earlier reports,\cite{Berkelbach_14a, Hestand_15a} the initially photogenerated singlet state, S$_1$, was found to be a lower Davydov component located at 1.85~eV (14\;900~cm$^{-1}$). The model furthermore identified the product triplet pair state, TT$_1$, located at 1.72~eV (13\;860~cm$^{-1}$), consisting of closely-spaced pairs of triplet excitations. Interestingly, a replica state was found at 1.89~eV (15\;240~cm$^{-1}$), located slightly above S$_1$, and separated from TT$_1$ by the vibrational quantum, $\omega_0$. This state was found to consist of triplet pair states dressed by a single vibration, and labeled TT$_1^*$.

In the second paper, hereafter Paper II,\cite{Tempelaar_17b} the vibronic, microscopic model was held to an additional stringent test by comparing simulations with measurements\cite{Bakulin_16a} of two-dimensional electronic spectra (2DES). A satisfactory agreement confirmed the credibility to the model, and opened the opportunity to analyze the optical transitions associated with the triplet pair state, in particular those involving higher-lying triplet excitons, TT$_n$ ($n>1$). In addressing a lack of consensus in the literature regarding the nature of such excitons,\cite{Hellner_72a, Ashpole_74a, Marciniak_07a, Pabst_08a, Marciniak_09a, Thorsmolle_09a, Thorsmolle_09b, Rao_10a, Smith_10a, Rao_11a, Wilson_13a, Smith_13a, Bakulin_16a, Khan_17b} we found that a single state with $n=2$ suffices to account for all features detected in 2DES experiments.

Building upon the work reported in Papers I\cite{Tempelaar_17a} and II,\cite{Tempelaar_17b} in the present article we combine our theoretical model with a dynamical framework in order to perform a study of the initial steps of the time evolution following photoexcitation in singlet fission, employing pentacene single crystals as an exemplary case. The dynamics is simulated using Redfield theory, perturbatively accounting for weakly-coupled Holstein modes \textit{other} than $\omega_0$. This approach enables us to evaluate singlet fission among multiple interacting molecules under conditions close to those in crystalline materials, and assess the effect of high-frequency Holstein modes. It also allows us to address the role of Peierls-type coupling, which is accounted for using the frozen modes formalism\cite{Montoya-Castillo_15a} which treats the low-frequency librational modes as a source of static disorder. In addition to both types of vibronic coupling, we identify a thermodynamic mechanism unique to crystalline environments as a driving force for singlet fission. A similar idea was previously proposed to explain experiments on tetracene,\cite{Chan_12a} and subsequently explored theoretically\cite{Kolomeisky_14a, Teichen_15a} albeit without non-perturbative vibronic coupling. We here show thermodynamics to drive the conversion between \textit{vibronically resonant} singlet and triplet states in crystalline pentacene at a time scale consistent with the experimentally determined fission rates.\cite{Chan_11a, Wilson_11a, Bakulin_16a}

This paper is organized as follows. We introduce the theoretical framework in Sec.~\ref{Sec_Theory}, briefly reviewing the model aspects introduced in Paper I,\cite{Tempelaar_17a} and emphasizing the model additions that allow for the dynamical simulations of singlet fission. We then continue to discuss our results in Sec.~\ref{Sec_Results}, paying special attention to the impact of the frequency $\omega_0$, the thermodynamic mechanisms induced by extended crystal environments, and Peierls coupling. We conclude in Sec.~\ref{Sec_Conclusions}.

\section{Theory}\label{Sec_Theory}

Since the theoretical model adopted in this work has in large part been described in Paper I,\cite{Tempelaar_17a} we keep its discussion brief, and focus mainly on the newly added components to the framework for the simulation of the fission dynamics. {We nonetheless start with a generic electronic-vibrational Hamiltonian, which helps to clarify the various approximations taken in our model.}

\subsection{Basis set and Hamiltonian}\label{Sec_Basis}

{The interactions among electronic and vibrational degrees of freedom are governed by the (total) Hamiltonian
\begin{align}
\Op H_\T{tot}=\Op H_\T{el}+\Op H_\T{vib}+\Op H_\T{el-vib},
\end{align}
with the electronic part given by
\begin{align}
\label{Eq_ElHamiltonian}
\Op H_\T{el}=\sum_i\Ket{i}\Bra{i}E_i+\sum_{i,j}\Ket{i}\Bra{j}V_{ij}.
\end{align}
Here, $i$ and $j$ run over the electronic basis states, and $E_i$ and $V_{ij}$ denote the basis state energies and interactions, respectively. As detailed in Paper I, our model describes crystalline fission materials using a diabatic basis involving for each molecule a set of electronic excitations including a singlet (s$_1$), triplet (t$_1$), cationic (c), and anionic (a) state, in addition to the (singlet) ground state (s$_0$).} The relevant electronic subspace of the crystal entails the manifold of single s$_1$ excitations and pairs of triplets and charged states, such that all basis states have zero spin and are charge neutral. Lower case is used for diabatic states in order to distinguish them from adiabatic states such as S$_1$ and TT$_1$. {We note that the utility of a diabatic basis (as a practical and physically motivated alternative to an adiabatic basis) to describe singlet fission is comprehensively discussed in Ref.~\citenum{Berkelbach_14a}.

In the total Hamiltonian the electronic part is added with the free vibrational Hamiltonian
\begin{align}
\Op H_\T{vib}=\sum_k\Op H_\T{vib}^{(k)}=\sum_k\Big[\frac{1}{2}\Op{p}^2_k+\frac{1}{2}\omega_k^2\Op{q}^2_k\Big],
\end{align}
and a part describing the bilinear coupling between electronic and vibrational degrees of freedom,
\begin{align}
\Op H_\T{el-vib}=\sum_k\Op H_\T{el-vib}^{(k)}=\sum_k\sum_{i,j}\Ket{i}\Bra{j}c_{k,ij}\Op q_k,
\end{align}
where $k$ labels the vibrational modes, and the} coupling constants are determined by the spectral density
\begin{align}
J_{ij}(\omega)=\frac{\pi}{2}\sum_k\frac{c^2_{k,ij}}{\omega_k}\delta(\omega-\omega_k).
\end{align}
The electronic-vibrational interaction Hamiltonian includes Holstein ($i=j$) as well as Peierls ($i\neq j$) couplings, which modulate the energies of the electronic diabatic states and their interaction terms, respectively. {(The term Peierls coupling is used here to refer to off-diagonal coupling in general.)

Fundamental to the theoretical model employed here and introduced in Paper I is the selection of a single Holstein-coupled mode for each molecule $m$, denoted $k_m$, which is treated on equal footing with the electronic degrees of freedom. The resulting vibronic quantum system is described by the Hamiltonian,
\begin{align}
\Op H_\T{sys}\equiv\Op H_\T{el}+\sum_m\Op H_\T{vib}^{(k_m)}+\sum_m\Op H_\T{el-vib}^{(k_m)},
\end{align}
while the associated mode frequency is taken to be molecule independent, and denoted $\omega_0$. The vibronic coupling terms appearing in $\Op H_\T{el-vib}^{(k_m)}$ can be expressed as $c_{k_m,ij}=\delta_{i,j}\delta_{m_i,m}\omega_0\lambda_i$, where $m_i$ refers to the molecule at which diabatic basis state $i$ is localized, the delta functions impose the Holstein-type character of the mode, and $\lambda_i$ relates to the Huang-Rhys factor associated with state $i$, $\lambda_i^2$. One then arrives at the Holstein Hamiltonian detailed in Paper I.\cite{Tempelaar_17a} To accommodate this Hamiltonian, the diabatic basis set includes in addition to the aforementioned electronic excitations the vibrational states associated with the mode $\omega_0$, which are described in terms of the eigenfunctions in the potentials associated with the molecular excitations s$_0$, s$_1$, t$_1$, c, and a, with corresponding state labels $\nu$, $\tilde{\nu}$, $\bar{\nu}$, $\nu_+$, and $\nu_-$, respectively.} Scaling of the resulting Hilbert space is kept manageable by restricting the vibronic basis to single and two-particle states, such that the total number of electronically excited and purely vibrationally excited ($\nu\geq1$) molecules does not exceed two.\cite{Philpott_71a, Spano_02a, Hestand_15a} Further truncations are applied to the spatial separation of pairs of triplets, and ionic states, as well as couples of singlet and purely vibrational excitations.\cite{Tempelaar_17a}

Coupling to the inter- and intramolecular modes other than that of frequency $\omega_0$ is numerically prohibitive to treat exactly, and instead demands for approximate methods. In Papers I\cite{Tempelaar_17a} and II,\cite{Tempelaar_17b} a phenomenological treatment was provided via the application of lineshape broadening in the linear absorption and 2DES simulations. Here, in order to account for their non-adiabatic effect on the quantum dynamics, { we employ two different physically motivated, approximate methods for the Holstein and Peierls-coupled vibrational modes, as described in the following.}

\subsection{Reduced density matrix dynamics}\label{Sec_Redfield}

As an approximate method to account for the Holstein-coupled vibrational modes we employ Markovian Redfield theory, which provides the reduced density matrix dynamics generated by the total Hamiltonian, $\Op H_\T{tot}$. In an extensive exploration tailored to Holstein modes in fission materials, Berkelbach \textit{et al.}\cite{Berkelbach_13a, Berkelbach_13b, Berkelbach_14a}~have shown this weak-coupling quantum master equation approach to provide both accuracy and numerical efficiency, and it has since found steady application in the simulation of singlet fission.\cite{Mirjani_14a, Bakulin_16a, Fujihashi_17a, Morrison_17a, Castellanos_17a} Here, we briefly reiterate the applied formalism, referring to Refs.~\citenum{Berkelbach_13a} and \citenum{May_1} for details. Within the secular approximation, the reduced density matrix of the system is propagated as\cite{May_1}
\begin{align}
\dot\rho_{\alpha\beta}=-i(\omega_\alpha-\omega_\beta)\rho_{\alpha\beta}+\sum_{\gamma,\delta}R_{\alpha\beta\gamma\delta}\rho_{\gamma\delta},
\label{Eq_Redfield}
\end{align}
where the indices label the adiabatic states and energies obtained by solving the eigenvalue equation $\Op H_\T{sys}\Ket{\alpha}=\omega_\alpha\Ket{\alpha}$ ($\hbar=1$ is applied throughout this work, similarly to Paper I\cite{Tempelaar_17a}). The Redfield tensor is given by
\begin{align}
R_{\alpha\beta\gamma\delta}=\Gamma^+_{\delta\beta\alpha\gamma}+\Gamma^-_{\delta\beta\alpha\gamma}-\delta_{\delta\beta}\sum_{\kappa}\Gamma^+_{\alpha\kappa\kappa\gamma}-\delta_{\alpha\gamma}\sum_{\kappa}\Gamma^-_{\delta\kappa\kappa\beta},
\end{align}
with
{
\begin{align}
\Gamma^+_{\alpha\beta\gamma\delta}=\int_0^\infty\diff\tau e^{-i\omega_{\gamma\delta}\tau}\sum_iC_i(\tau)K_{\alpha,\beta}^{(i)}K_{\gamma,\delta}^{(i)},
\end{align}
and
\begin{align}
\Gamma^-_{\alpha\beta\gamma\delta}=\int_0^\infty\diff\tau e^{-i\omega_{\alpha\beta}\tau}\sum_iC_i^*(\tau)K_{\alpha,\beta}^{(i)}K_{\gamma,\delta}^{(i)},
\end{align}
Here, $K_{\alpha,\beta}^{(i)}\equiv\Bra{\alpha}i\rangle\langle i\Ket{\beta}$, and $C_{i}(t)$ is the thermal correlation function,
\begin{align}
C_i(t)=\frac{1}{\pi}\int_0^\infty\diff\omega\tilde J_{ii}(\omega)\Big[\coth\Big(\frac{\beta\omega}{2}\Big)\cos(\omega t)-i\sin(\omega t)\Big],
\end{align}
with $\beta=1/k_\T{B}T$ as the inverse temperature, and $\tilde J_{ii}(\omega)$ as the spectral density minus the contribution at $\omega_0$ (to avoid double counting). Note that $K_{\alpha,\beta}^{(i)}$ contains the purely electronic operator $\Ket{i}\Bra{i}$, as a result of which the bath induced fluctuations of different vibrational levels associated with the same electronic diabatic excitation are taken to be fully correlated. Such approach is taken for numerical convenience and to limit the number of parameters, although the degree of such correlation is expected to be limited in realistic materials (in particular in the form of vibrational dephasing), effects of which are worthy of investigation in future studies.}

Despite Redfield theory being inexpensive compared to alternative dynamical methods, the cost of integrating Eq.~\ref{Eq_Redfield} still scales as $N^4$, with $N$ the total number of basis states, which, in combination with the applied vibronic basis set, results in a rapidly growing computational expense with increasing number of molecules. On the other hand, most of the dynamics of interest in singlet fission takes place within a relatively narrow region on the low-energy side of the manifold of excited states (see Paper I\cite{Tempelaar_17a}). In our calculations, we utilize this observation by limiting the Redfield tensor to the $N'$ ($N'<N$) lowest-energy adiabatic states, while strategically choosing $N'$ so as to allow for the treatment of relatively large clusters with reasonable accuracy.

As illustrated in Fig.~\ref{Fig_Scheme}, singlet fission is initiated by photon absorption from the vibrationless ground state (vacuum state) of the system, labeled S$_0$. Neglecting polarization contributions, the dipole operator describing this process is given by
\begin{align}
\Op{M}=\mu\sum_m\Ket{\BS{\T{s}_0}_m}\Bra{\BS{\T{s}_1}_m}+\T{H.c.},
\end{align}
where the summand is a purely electronic operator which allows for transitions between the singlet ground and excited states at molecule $m$. Hence, photon absorption creates the initial state $\Ket{\Psi_\T{i}}=\Op{M}\Ket{\T{S}_0}$. This state commonly has a large overlap with one of the adiabatic eigenstates of $\Op{H}_\T{sys}$, which is referred to as S$_1$. In the Redfield calculations, we initiated the reduced density matrix as $\rho=\Ket{\T{S}_1}\Bra{\T{S}_1}$, while identifying S$_1$ by scanning the $N'$ lowest-energy eigenstates for a maximized overlap, $\vert\langle\alpha\Ket{\Psi_\T{i}}\vert^2$. Although this initiation condition is a proxy for an actual absorption event involving a polarized photon, previous calculations have shown the fission dynamics to be relatively insensitive to the specific conditions used.\cite{Berkelbach_14a}

\subsection{Frozen modes approach}\label{Sec_Arrested}

Peierls coupling was originally suggested to have a negligible effect on the fission dynamics based on Redfield theory arguments.\cite{Berkelbach_13b} In contrast, by employing a partial-linearized density matrix path integral formalism, Castellanos and Huo showed that such coupling can indeed impact singlet fission, highlighting that Redfield theory may fall short in describing this effect. Indeed, the relevant inter-molecular vibrations have relatively low frequencies and high reorganization energies, ingredients of strong coupling, highly non-Markovian dynamics that falls outside the realm of applicability of Redfield theory. However, when Redfield calculations based on other (Markovian) modes are performed, such strong coupling and non-Markovian effects can be included by application of the frozen modes approach.\cite{Montoya-Castillo_15a} Accordingly, the lowest frequency modes are arrested and included as static disorder in the system Hamiltonian. This treatment has been demonstrated to result in a quantitative description of dynamics by proof-of-principles calculations of a spin-boson system and a multi-site photosynthetic complex,\cite{Montoya-Castillo_15a} {while its applicability was later confirmed\cite{Castellanos_17a} for Peierls coupling in fission materials, for which the associated time scale has been calculated to be on the order of 1 ps.\cite{Troisi_06a}}

Following the frozen mode approach, we model the Peierls coupling as static random modulations of the charge overlap integrals appearing in the system Hamiltonian, $\hat{H}_{\T{sys}}$ (see Paper I\cite{Tempelaar_17a} for details). Accordingly, we implement this coupling by adding each charge overlap integral with a random contribution such that for the HOMO-HOMO integral between molecules $m$ and $m'$
\begin{align}
t_{m,m'}^\T{HH}=\langle t_{m,m'}^\T{HH}\rangle+\delta t_{m,m'}^\T{HH},
\end{align}
and analogous for the LUMO-LUMO and HOMO-LUMO integrals. Here, $\langle t_{m,m'}^\T{HH}\rangle$ is the average integral value (discussed in Paper I\cite{Tempelaar_17a}), and $\delta t_{m,m'}^\T{HH}$ is drawn from a random distribution. The total set of random contributions to $\hat{H}_{\T{sys}}$ is taken to be uncorrelated. It is noteworthy that $\delta t_{m,m'}^\T{HH}=\delta t_{m',m}^\T{HH}$ has to be fulfilled for physical reasons, in contrast to $\delta t_{m,m'}^\T{HL}$ and $\delta t_{m',m}^\T{HL}$ which correspond to different electron transfer mechanisms.

\subsection{Parameters for pentacene}\label{Sec_Parms}

Paper I\cite{Tempelaar_17a} extensively discussed the parametrization of the system Hamiltonian, $\hat{H}_{\T{sys}}$, for pentacene single crystals. Besides the addition of the parameters involved with the bath, the present work leaves the parametrization fully unaltered, with the exception of the Holstein mode of frequency $\omega_0$ incorporated in the quantum basis set. In Paper I\cite{Tempelaar_17a} this mode was parametrized phenomenologically by fitting to the high-frequency vibronic progression observed in absorption spectroscopy of pentacene monomers.\cite{Yamagata_11a, Beljonne_13a, Berkelbach_13b, Hestand_15a} The obtained frequency of 1380~cm$^{-1}$ has been ascribed to a symmetric ring-stretching mode,\cite{Yamagata_11a, Hestand_15a} found to be common to $\pi$-conjugated molecules.\cite{Spano_10a} However, Raman spectroscopic experiments\cite{Brillante_02a} and first principles calculations\cite{Girlando_11a, Mirjani_14a} have identified three distinct sub-bands located roughly at 1150 cm$^{-1}$, 1400 cm$^{-1}$, and 1600 cm$^{-1}$, all with similar coupling strengths, with additional attributions to CH bending and CC bending character. We therefore supplement the calculations based on the original ($\omega_0=1380$~cm$^{-1}$) vibronic model with those in which $\omega_0$ is switched between the vibrational sub-bands. In such cases, the corresponding Huang-Rhys (HR) factors ($\lambda_{\T{s}_1}^2$, $\lambda_{\T{a}}^2$, $\lambda_{\T{c}}^2$, and $\lambda_{\T{t}_1}^2$ -- see Paper I\cite{Tempelaar_17a}) are taken to be 1/3 times the values from the original model.\cite{Girlando_11a}

Besides the aforementioned high-frequency vibrations, a variety of Holstein modes at lower frequencies has been observed for pentacene, in particular in the region from 50 to 300 cm$^{-1}$.\cite{Brillante_02a, Girlando_11a, Mirjani_14a} In order to account for this Raman activity, we take the Holstein spectral density to be of Debye form (Ohmic form with Lorentzian cutoff),
\begin{align}
\tilde J_{ii}(\omega)=2\lambda\Omega\frac{\omega}{\omega^2+\Omega^2},
\end{align}
using reasonable values for the associated parameters, that is, a characteristic frequency $\Omega=200$~cm$^{-1}$ and a reorganization energy $\lambda=100$~cm$^{-1}$. {(The contribution at $\omega_0$ is not rigorously discarded in a spectral density of this form, but is negligible at the relevant frequency range, so that the effects of double counting are insignificant.)} We note that it would be interesting to explore larger values of $\lambda$,\cite{Ito_15a} which we expect to enhance the fission rate, but which lies outside the realm of validity of the Redfield formalism employed in the present article.\cite{Montoya-Castillo_15a} In our Redfield calculations, {the temperature is fixed at 300~K, and a time step is applied} of 1~fs, which is sufficient to reach convergence.

In parametrizing the Peierls modes we have followed the example of Castellanos and Huo\cite{Castellanos_17a} who applied a Gaussian distribution with a standard deviation $\sigma=256$~cm$^{-1}$ to all charge overlap integrals for the strongest-coupled pentacene pair in the crystal geometry. This number is in reasonable agreement with {room temperature} atomistic simulations of the Peierls activity in crystalline pentacene.\cite{Troisi_06a, Girlando_11a} Since we are dealing with electronic interactions beyond a single pentacene pair, we have determined for each integral the \textit{relative} standard deviation, $\bar{\sigma}=\sigma/\langle t\rangle$, with $\langle t\rangle$ taken to be the Hartree-Fock calculated integral between the two pentacenes considered in Ref.~\citenum{Castellanos_17a}. This yields $\bar{\sigma}^\T{HH}=0.22$, $\bar{\sigma}^\T{LL}=0.27$, and $\bar{\sigma}^\T{HL}=0.22$. Every charge overlap integral in the crystal is then drawn from a normal distribution such that for the HOMO-HOMO integral between molecules $m$ and $m'$,
\begin{align}
P(\delta t_{m,m'}^\T{HH})=(2\pi\sigma_{m,m'}^\T{HH})^{-1/2}\exp(-\delta{t_{m,m'}^\T{HH}}^2/2{\sigma_{m,m'}^\T{HH}}^2),\nonumber
\end{align}
with the \textit{absolute} standard deviation $\sigma_{m,m'}^\T{HH}=\bar{\sigma}^\T{HH}\langle t_{m,m'}^\T{HH}\rangle$, and analogously for the LUMO-LUMO and HOMO-LUMO integrals. We found the resulting dynamics to converge rapidly with increasing number of disordered samples, and 500 realizations to be sufficient to obtain converged results.

In Papers I\cite{Tempelaar_17a} and II\cite{Tempelaar_17b} crystalline pentacene was considered as a quasi-two-dimensional material extending in the crystallographic $ab$ plane, owing to the weak interactions in the $c$ direction.\cite{Yamagata_11a} The combination of the Redfield propagation scheme and the applied vibronic basis set results in rapidly diverging computational expenses with increasing size of the pentacene crystal in the present article. For that reason we are limited to supercells consisting of $2\times2$ unit cell configurations with 2 molecules each, corresponds to a cluster of 8 molecules. Periodic boundaries are imposed in order to limit the finite size effects, and to ensure the cluster adheres to the symmetry of pristine pentacene crystals.\cite{Teichen_15a, Refaely-Abramson_17a} While we find $2\times2$ supercells sufficient to qualitatively study singlet fission in crystalline pentacene, our results in Paper I\cite{Tempelaar_17a} indicated that the adiabatic energy associated with S$_1$ only converges for $3\times3$ supercells and beyond. To avoid spurious results, we have adjusted the diabatic singlet and triplet energies, $E_{\T{s}_1}$ and $E_{\T{t}_1\T{t}_1}$, such that for each of the applied crystal sizes the energy of S$_1$ coincides with its converged value shown in Fig.~\ref{Fig_Scheme} (see Paper I\cite{Tempelaar_17a} for an evaluation of the S$_1$ energy, as well as a discussion of $E_{\T{s}_1}$ and $E_{\T{t}_1\T{t}_1}$). We note that this procedure has in essence been followed by studies on singlet fission using few-state models as proxies for a extended crystals.\cite{Yost_14a, Bakulin_16a, Monahan_16a, Fujihashi_17a, Morrison_17a, Sun_17a} Lastly, a truncation of the Redfield tensor (see Sec.~\ref{Sec_Redfield}) is performed such that adiabatic states having energies higher than 1.92~eV (15\;500~cm$^{-1}$) are excluded. A comparison with non-truncated Redfield calculations is shown in Fig.~\ref{Fig_Dynamics}(b) and (c) for a $1\times1$ supercell with $\omega_0=1150$~cm$^{-1}$ excluding and including Peierls coupling, respectively, illustrating that high accuracy is retained in both cases.

\section{Results and discussion}\label{Sec_Results}

\subsection{Vibronic resonance}\label{Sec_Resonance}

\begin{figure}
\includegraphics{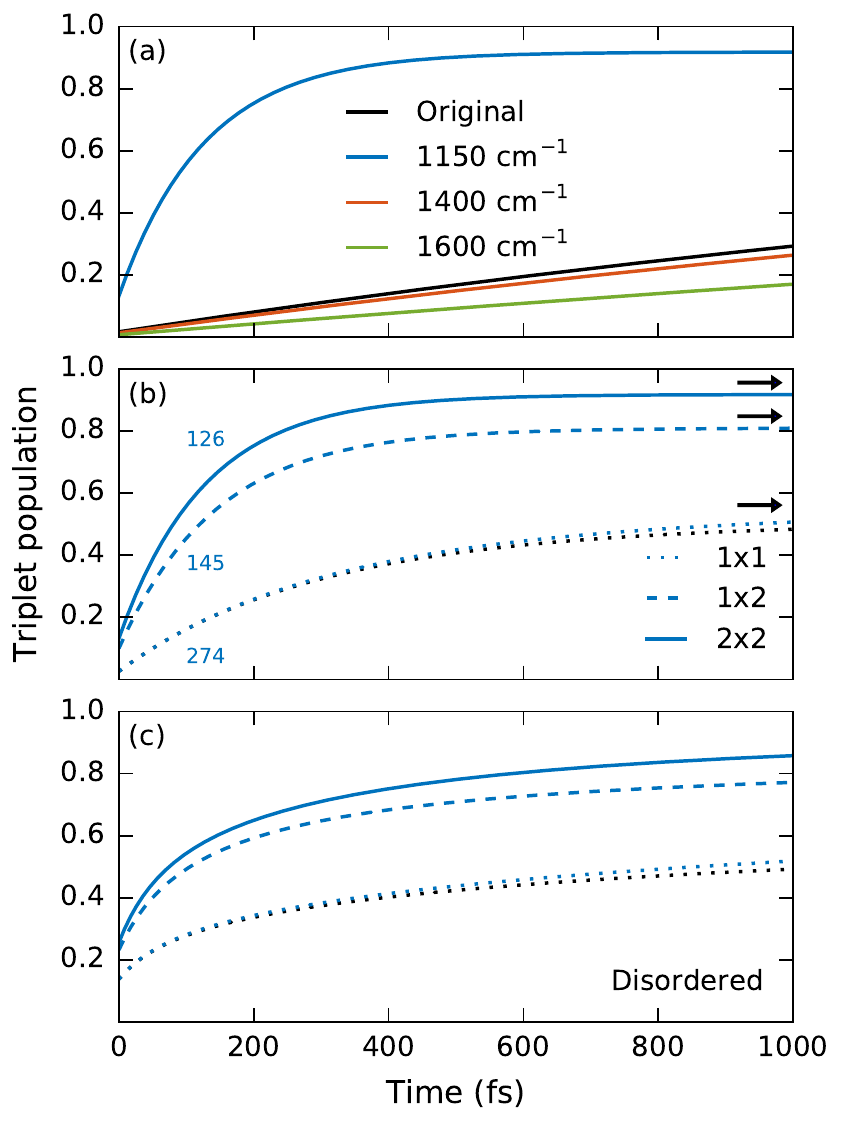}
\caption{Singlet fission dynamics of a $2\times2$ pentacene supercell generated by the different vibronic models (a), and for different supercell sizes in the absence (b) and presence (c) of Peierls-type disorder in the charge overlap integrals. Curves indicate the total population of diabatic triplet pair states as a function of time, resulting from the original vibronic model (black), and with $\omega_0=1150$~cm$^{-1}$ (blue), 1400~cm$^{-1}$ (red), and 1600~cm$^{-1}$ (green) -- see Sec.~\ref{Sec_Parms} for details. Dotted, dashed, and solid lines correspond to $1\times 1$, $1\times 2$, and $2\times 2$ supercells, respectively. For each curve in panel (b), the associated number indicates the time constant (in fs) resulting from an exponential fit, whereas the black arrow denotes the equilibrium population based on a Boltzmann distribution involving S$_1$ and TT$_1^*$ -- see Eq.~\ref{Eq_Boltzmann}. Black dotted curves represent the $1\times 1$ supercell dynamics obtained without truncating the Redfield tensor -- see Sec.~\ref{Sec_Redfield}.}
\label{Fig_Dynamics}
\end{figure}

We begin by presenting results from Redfield calculations in which the contribution from the Peierls coupling is discarded, following the example of virtually all previous theoretical studies on singlet fission dynamics, and reserve a thorough discussion of its effect on the dynamics for Sec.~\ref{Sec_Peierls}. In Fig.~\ref{Fig_Dynamics}(a), we compare the fission dynamics resulting from the different vibronic models applied to a $2\times2$ supercell. Shown is the total population of diabatic triplet pair states ($\T{t}_1\T{t}_1$) as a function of time. Interestingly, the original vibronic model with $\omega_0=1380$~cm$^{-1}$ (detailed in Paper I\cite{Tempelaar_17a}) is found to result in a fission time constant that is an order of magnitude \textit{longer} than the 270~fs obtained by Berkelbach \textit{et al.}~based on a purely electronic model.\cite{Berkelbach_14a} This might at first come as a surprise, as the non-perturbative inclusion of Raman active modes is thought to enhance the fission rate.\cite{Bakulin_16a, Monahan_16a, Fujihashi_17a, Morrison_17a, Huang_17a} However, our results show this enhancement to be remarkably sensitive to the mode frequency, $\omega_0$, as also suggested recently based on phenomenological modeling.\cite{Fujihashi_17a} This is borne out when considering the vibronic models where $\omega_0$ is switched between the three vibrational sub-bands underlying the 1380~cm$^{-1}$ progression in pentacene (we concomitantly divide the associated HR factors by a factor of 3 to account for the coupling strength of these sub-bands -- see Sec.~\ref{Sec_Parms}). Whereas the sub-bands at 1400~cm$^{-1}$ and 1600~cm$^{-1}$ result in even lower fission rates, setting $\omega_0=1150$~cm$^{-1}$ yields a rate that is \textit{faster} than that by Berkelbach \textit{et al.}, and well within the range of experimental measurements for pentacene crystals.\cite{Chan_11a, Wilson_11a, Bakulin_16a} As such, our results identify the 1150~cm$^{-1}$ mode to be a key facilitator for singlet fission in bulk pentacene.

\begin{figure}
\includegraphics{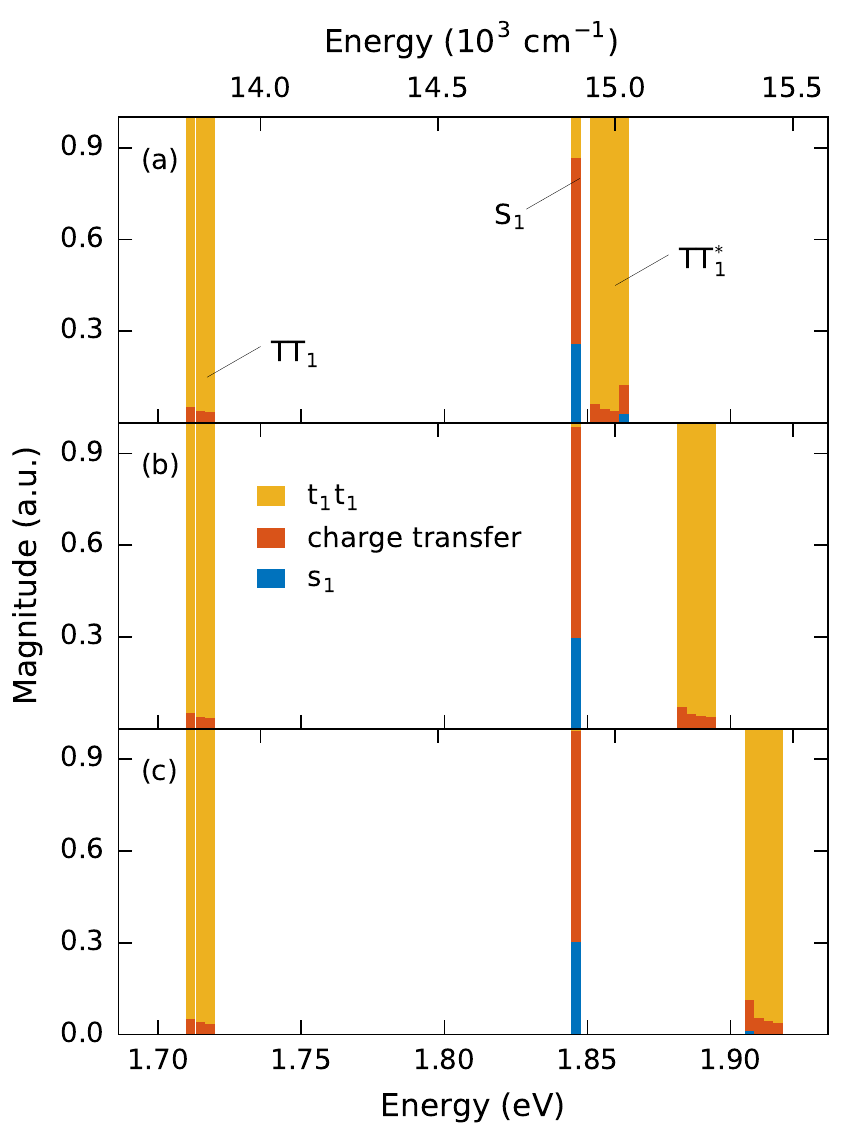}
\caption{Bar plot labeling for each adiabatic state the composition of diabatic singlet (blue), charge transfer (red), and triplet states (yellow), for the vibronics model with $\omega_0=1150$~cm$^{-1}$ (a), 1400~cm$^{-1}$ (b), and 1600~cm$^{-1}$. Shown are all adiabatic states with finite (nonzero) oscillator strength relative to the ground state (see Paper I\cite{Tempelaar_17a}). States in panel (a) are labeled in accordance with their diabatic content, and consistent with Fig.~\ref{Fig_Scheme}. Note that TT$_1$ and TT$_1^*$ are represented by groups of states, whereas S$_1$ consists of a single state.}
\label{Fig_Sticks}
\end{figure}

The strong sensitivity of the fission rate to $\omega_0$ is rationalized by characterizing the adiabatic states in pentacene crystals in terms of their electronic composition and energy. Shown in Fig.~\ref{Fig_Sticks} is such a characterization involving all states with finite oscillator strength for the vibronic models associated with each of the sub-bands, analogous to Fig.~2(c) of Paper I.\cite{Tempelaar_17a} Common to all sub-bands, and consistent with Fig.~\ref{Fig_Scheme}, a group of states of predominantly triplet composition is found at 1.72~eV, which is identified as the correlated triplet pair, TT$_1$, while a single state residing at 1.85~eV carries the bulk of the singlet admixture, which is accordingly labeled S$_1$. A replica of TT$_1$ is found at higher energies, consisting of triplet pairs dressed with one vibrational quantum, labeled TT$_1^*$. Comparing the different vibronic models, TT$_1^*$ can be seen to have varying degrees of resonance with S$_1$ depending on $\omega_0$, which correlates with the associated fission rate from Fig.~\ref{Fig_Dynamics}(a), and which is maximized for $\omega_0=1150$~cm$^{-1}$.

To explain why the fission rates resulting from our vibronic models can be both lower and higher than the purely electronic calculations by Berkelbach \textit{et al.}, we note that in the latter case the high-frequency modes under consideration were taken to be part of a continuous spectral density, taken to be of Debye form, peaked at 1450~cm$^{-1}$.\cite{Berkelbach_14a} As such, all vibrational sub-bands and their possible beneficial effect on singlet fission were incorporated, albeit perturbatively. Our vibronic models provide a non-perturbative description of the sub-bands, but is restricted to one discrete mode at a time. Hence, with an exact treatment of the 1150~cm$^{-1}$ mode we find comparatively faster fission, while the omission of this mode results in slower dynamics.

\subsection{Thermodynamics}\label{Sec_Thermo}

It is important to note that vibronic resonance between singlet and triplet pair states \textit{by itself} is insufficient to account for rapid singlet fission in crystalline pentacene. Instead, our results show that the degree at which this resonance impacts the fission dynamics depends critically on the applied crystal dimensions. This is demonstrated in Fig.~\ref{Fig_Dynamics}(b), where the time-dependent populations of diabatic triplet pair states are compared for $1\times1$, $1\times2$, and $2\times2$ supercells, with $\omega_0=1150$~cm$^{-1}$ ($1\times2$ meaning two unit cells extending in the crystallographic $b$ direction, along which we found size effects are most pronounced). The dynamics is found to be highly exponential in all cases, so that the associated time constants are readily obtained through exponential fittings. The results are indicated in Fig.~\ref{Fig_Dynamics}(b), and can be seen to decrease with increasing supercell size, from 274~fs for a $1\times1$ supercell to 126~fs for a $2\times2$ supercell. Interestingly, we find a concomitant increase in the apparent asymptotic triplet populations.

\begin{figure}
\includegraphics{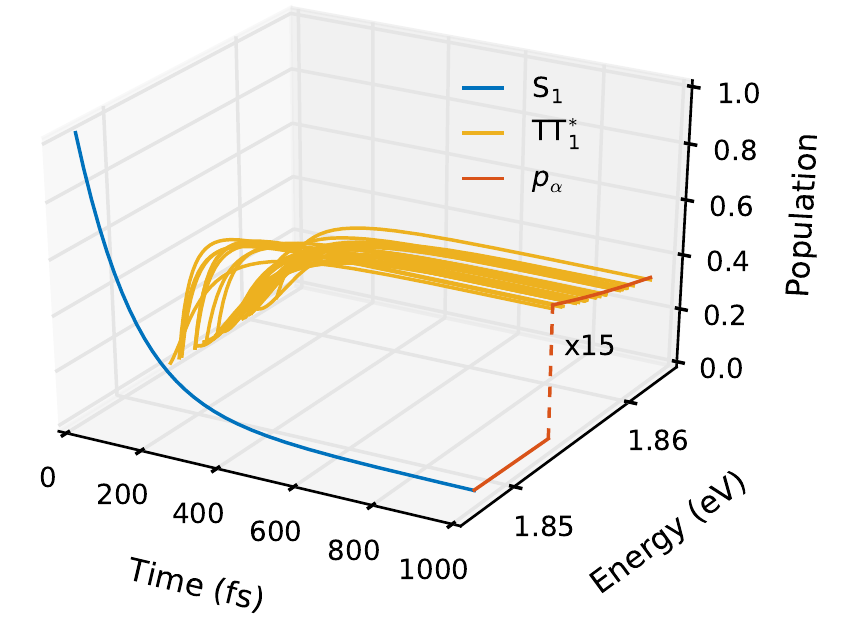}
\caption{Singlet fission dynamics dissected into individual adiabatic states. Shown are the time-dependent populations of S$_1$ (blue) and the adiabatic states underlying TT$_1^*$ (yellow, enhanced by a factor of 15 for the ease of demonstration), for a $2\times2$ supercell with $\omega_0=1150$~cm$^{-1}$. Red curve represents the Boltzmann distribution of the collection of S$_1$ and the TT$_1^*$ states -- see Eq.~\ref{Eq_Boltzmann}.}
\label{Fig_Detail}
\end{figure}

As it turns out, the above observations are related to an interesting aspect of the excited states shown in Fig.~\ref{Fig_Sticks}, namely that while incident light couples predominantly to only a single singlet-dominated adiabatic state, multiple triplet-dominated states are available to become subsequently populated. On one hand, this difference in numbers emerges from the bi-exciton character of the triplet pairs, which in the purely-electronic representation results in $M(M-1)/2$ combinations given an ensemble of $M$ molecules, in contrast to a total of $M$ singlet excitons. Perhaps more significant, however, is the contrast in collective characteristics between singlet and triplet excitons.\cite{Teichen_15a} Whereas triplets remain largely localized, singlet states tend to delocalize over a considerable number of molecules in crystalline environments.\cite{Cudazzo_12a} For pentacene, the resulting Davydov splitting induces an energetic separation between singlet states, while most of the oscillator strength is concentrated in the lower Davydov component, S$_1$.\cite{Yamagata_11a, Beljonne_13a, Berkelbach_14a, Hestand_15a, Tempelaar_17a} This sharpens the contrast between the number of singlet and triplet states involved in the fission process. To demonstrate this, we dissect in Fig.~\ref{Fig_Detail} the dynamics for the $2\times2$ supercell with $\omega_0=1150$~cm$^{-1}$ into individual adiabatic states. About thirty states, representative of TT$_1^*$, can be seen to acquire population within 200~fs, and although their individual population never exceeds a value of 0.05, their accumulative effect is found to rapidly deplete S$_1$. Furthermore, as evinced by Fig.~\ref{Fig_Detail}, within the time frame of 1 ps the asymptotic populations can largely be understood based on a Boltzmann distribution involving the states underlying S$_1$ and TT$_1^*$,
\begin{align}
\label{Eq_Boltzmann}
p_\alpha=\frac{1}{Z}e^{-\omega_\alpha\beta},
\end{align}
with $Z=\sum_{\alpha\in\T{S}_1,\T{TT}_1^*}\exp(-\omega_\alpha\beta)$. It is worth noting that population transfer towards the band-bottom states underlying TT$_1$ is expected to occur at longer time scales, but this effect is found to be small within the time frame under consideration.

\begin{figure}
\includegraphics{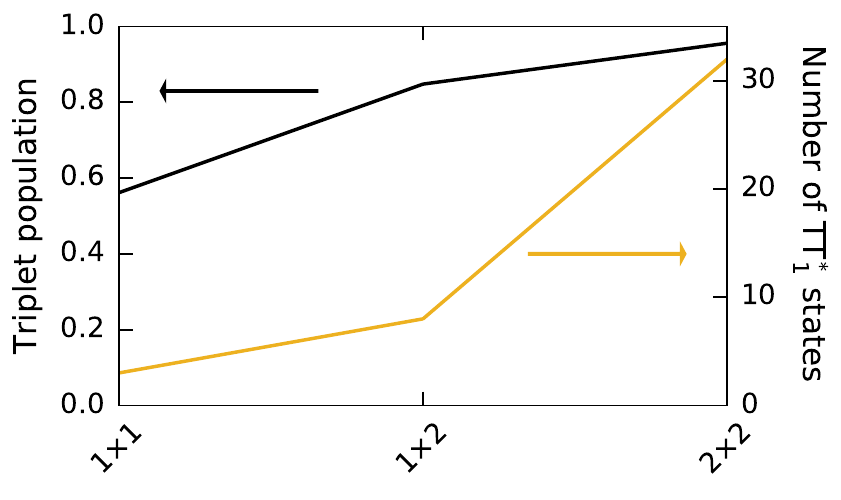}
\caption{Asymptotic triplet population (black) and total number of states underlying TT$_1^*$ (yellow) as a function of the supercell size, resulting from the vibronic model with $\omega_0=1150$~cm$^{-1}$.}
\label{Fig_Sizes}
\end{figure}

Fig.~\ref{Fig_Sizes} depicts the total number of adiabatic states underlying TT$_1^*$ as a function of the supercell size, together with the total triplet population predicted by Eq.~\ref{Eq_Boltzmann}. The latter is seen to monotonically increase with expanding supercell, and by plotting these data as arrows alongside the associated transients in Fig.~\ref{Fig_Dynamics}(b) we find that the asymptotic behavior of triplets is indeed well accounted for by the Boltzmann distribution for all supercell sizes. Moreover, Fig.~\ref{Fig_Sizes} shows a concerted, rapid build up of the number of TT$_1^*$ states with increasing supercell. Note that this number deviates from $M(M-1)/2$ due to vibronic mixing and a truncation of the triplet-triplet separation, see Sec.~\ref{Sec_Basis}. In accordance with Fig.~\ref{Fig_Dynamics}(b), this build up not only correlates with an increase of the asymptotic triplet population, but also with an enhancement in the fission rate.

Hence, in addition to the vibronic resonance criterion, we identify thermodynamic counting to be of key importance to the rate of singlet fission in pentacene crystals, namely that an increase in entropy associated with the transfer from a single singlet to a multitude of triplets drives the conversion process. It should be noticed that although not explicitly discussed, such thermodynamic arguments are operative in the older model of fission in pentacene crystals of Ref.~\citenum{Berkelbach_14a}. Furthermore, some works have proposed a similar principle based on entropy to explain the efficacy of singlet fission in tetracene crystals,\cite{Chan_12a, Kolomeisky_14a} in which the conversion from S$_1$ to TT$_1$ is moderately endothermic. Interestingly, fission in pentacene is \textit{in principle} exothermic, but the conversion from the adiabatic state S$_1$ to TT$_1^*$ is slightly endothermic. Hence, distinct from tetracene, we find thermodynamics to act alongside of vibronic resonance in crystalline pentacene to produce an ultra-fast $\sim$100~fs fission time constant.

\subsection{Peierls coupling}\label{Sec_Peierls}

Although Peierls coupling is commonly discarded in theoretical investigations of fission dynamics, a recent study by Castellanos and Huo suggested this factor to be of significance.\cite{Castellanos_17a} The rationale for this finds its origin in an effective destructive interference between two pathways through which neighboring pentacenes in the crystal environment can convert a singlet into two triplets, induced by the off-diagonal elements of the {electronic Hamiltonian, $H_\T{el}$ (Eq.~\ref{Eq_ElHamiltonian}).} Peierls coupling, which modulates these elements, can partially break the interference, which possibly affects the fission rate.\cite{Castellanos_17a} Following this idea, we proceed to evaluate the dynamics predicted by our model in which such Peierls modulations are considered as static disorder in $H_\T{sys}$, employing the frozen modes approach described in Sec.~\ref{Sec_Arrested}.

Results for $\omega_0=1150$~cm$^{-1}$, shown in Fig.~\ref{Fig_Dynamics}(c) for different supercell sizes, confirm that Peierls-type disorder indeed impacts the fission dynamics. In contrast to the disorder-free case, the triplet populations are found to behave in a non-exponential manner, complicating the extraction of a single time constant for each of the population curves {(the static Peierls modes result in a different time scale for each disordered crystal, adding up to form such non-exponential ensemble dynamics)}. However, visual inspection shows that the Peierls disorder slows down the fission dynamics somewhat. This forms a surprising contrast with Castellanos and Huo,\cite{Castellanos_17a} where Peierls coupling was found to \textit{enhance} the fission rate. We attribute this difference to the synergistic effects of vibronic resonance and thermodynamics (see Secs.~\ref{Sec_Resonance} and \ref{Sec_Thermo}) included in our model, both of which were not considered by Castellanos and Huo.\cite{Castellanos_17a} In the disorder-free case, these effects allow for rapid fission in spite of the destructively interfering pathways in $H_\T{sys}$. Inclusion of Peierls disorder not only attenuates this interference, but also affects the eigenenergies of $H_\T{sys}$. This breaks the resonance condition between S$_1$ and TT$_1^*$, which in turn inhibits the fission rate. As such, Peierls coupling can be both beneficial and detrimental to singlet fission.

We should, however, add three important remarks with regards to the reported Peierls disordered calculations. First, an aspect of Peierls coupling that is missing in the frozen modes approach is the time-dependent fluctuations of the charge-overlap integrals which, although slow, can significantly alter the off-diagonal elements of $H_\T{sys}$ over the course of the fission process.\cite{Troisi_06a} These fluctuations dynamically bring S$_1$ and TT$_1^*$ in and out of resonance, and the instants where resonance holds will be characterized by a high fission rate, in accordance with our findings in Secs.~\ref{Sec_Resonance} and \ref{Sec_Thermo}. Second, a recent study wherein pentacene thin films were partly substituted with non-interacting spacer molecules showed an invariance of the fission rate to the degree of substituted impurities,\cite{Broch_18a} suggesting that the photoexcited singlet state rapidly finds local regions of pristine crystallinity (``hot spots'') through which ultra-fast fission proceeds. It is likely that a similar effect, not captured by the limited crystal sizes applied in our work, mitigates the detrimental effects of Peierls coupling. Accordingly, the singlet exciton migrates to hot spots where vibronic resonance between S$_1$ and TT$_1^*$ is maintained, following which immediate fission occurs through thermodynamics.

As a last remark, since Peierls coupling affects the eigenvalue equation involving $H_\T{sys}$, we anticipate its effect to extend over linear absorption and 2DES through their dependence on the eigenenergies and eigenvectors. The parametrization of the vibronic exciton model employed in this article, as well as Papers I\cite{Tempelaar_17a} and II,\cite{Tempelaar_17b} has been validated based on these spectroscopic techniques,\cite{Yamagata_11a, Hestand_15a, Tempelaar_17a, Tempelaar_17b} but without the inclusion of Peierls modes. It is therefore likely that some aspects of these modes have been effectively accounted for through other model degrees of freedom, which are double counted once Peierls disorder is added to the model. All considerations brought up here are suggestive of an enhancement of the fission rate compared to the results shown in Fig.~\ref{Fig_Dynamics}(c), and it is even conceivable that Peierls coupling acts alongside vibronic resonance and thermodynamics to promote rapid singlet fission in crystalline pentacene depending on the material details. A rigorous approach to establish this would be to re-parametrize the vibronic exciton model starting with Peierls modes as an elementary component. This program forms a significant effort and we consider this a fruitful future direction for microscopic theories of singlet fission.

\section{Conclusions}\label{Sec_Conclusions}

To summarize, we have extended the previously developed vibronic exciton theory to dynamically study singlet fission under realistic crystalline conditions, with a concrete focus on the ultra-fast fission rate in pentacene single crystals as a representative problem that challenges our general understanding of the process. Our theory involves a microscopic set of molecular degrees of freedom, including for each molecule a single vibrational mode of frequency $\omega_0$. This enabled us to provide an explicit quantum treatment of the Holstein-type vibronic coupling involving a high-frequency intramolecular vibration, which has recently been suggested to profoundly impact singlet fission. A perturbative description of weakly-coupled Holstein modes of distinct frequency scales is invoked through Redfield theory, while the frozen modes approach allowed for the inclusion of Peierls-type vibronic coupling.

Our numerical results identified three factors impacting the fission rate in crystalline pentacene. Firstly, the rate was found to be remarkably sensitive to the frequency $\omega_0$, which modulates the resonance conditions between the photo-prepared singlet state and the vibrationally-dressed triplet pair product states. Whereas we previously parametrized $\omega_0$ phenomenologically in order for it to represent an effective progression-building mode observable in linear absorption and 2DES, here we alternated $\omega_0$ among the sub-bands known to underly this progression. Following this procedure we identified the band located around 1150~cm$^{-1}$ as the key facilitator of rapid fission in solid pentacene. Alongside this vibronic resonance criterion, our calculations demonstrated that thermodynamics drives the conversion from the initial singlet exciton into a multitude of vibrationally-dressed triplet pairs, { collectively labeled as TT$^*$}. Lastly, we addressed a recent report indicating the importance of Peierls coupling. In contrast to that report, we find that such coupling does not enhance the fission rate \textit{per se}, but that its effect on singlet fission depends sensitively on the synergistic effects of vibronic resonance and thermodynamics.

{In this work we have focused on fission in solid pentacene at room temperature, with the primary aim to interpret ultra-fast time-resolved spectroscopic experiments taken under such conditions.\cite{Chan_11a, Wilson_11a, Bakulin_16a} Based on our findings, a certain degree of temperature dependence is expected for this process, in particular with regards to the endothermic nature of the rapid conversion from S$_1$ to TT$_1^*$ found in our calculations. Indeed, repeating the (Peierls disorder-free) simulations from Fig.~\ref{Fig_Dynamics}(b) at temperatures of 200~K and 100~K yields a rate amounting to roughly 60\% and 25\%, respectively, of the rate at 300~K, although the relative enhancement with increasing crystal size is retained at all temperatures. However, a drop in temperature is likely to simultaneously enhance the delocalization length and mobility of the singlet excitation, through which ``hot spots'' with favorable fission conditions can be accessed more rapidly (as discussed in Sec.~\ref{Sec_Peierls}), which could (partly) compensate for the temperature trend associated with the thermodynamically driven endothermic step. A further temperature dependence is associated with the low-frequency Peierls-coupled modes whose fluctuations are expected to weaken upon cooling. Ultimately, fission in pentacene populates the (vibrationally bare) TT state, to which the transition from S$_1$ is markedly downhill in energy. This is consistent with the lack of temperature dependence of the singlet to triplet conversion yield found for pentacene in measurements taken at 200~ps after photoexcitation;\cite{Thorsmolle_09a} a timescale at which this downhill energy transfer has fully completed. However, given the nontrivial interplay of the temperature dependent factors highlighted in this work, temperature resolved spectroscopic measurements of singlet fission in pentacene at ultra-fast time scales could be very instructive.}

The unfavorable scaling of the applied vibronic basis with the number of molecules poses a firm numerical constraint on the crystal sizes reachable in the dynamical calculations presented here. Still, our approach afforded flexibility to identify a pronounced increase of the fission rate with expanding supercells. This trend is consistent with the aforementioned thermodynamic mechanism becoming more effective with increasing number of molecules, and our results in Fig.~\ref{Fig_Dynamics}(b) suggested this principle to drive sub-100 fs fission in extended crystalline environments. As such, expanding the supercell sizes applied in the present work appears as an obvious option to improve the model. Retaining a quantum description of the high-frequency Holstein-mode poses a considerable methodological challenge, but resolving this would open up the possibility to microscopically study the spatial separation of triplets\cite{Pensack_16a, Breen_17a} as an additional benefit. {Irrespective of such practical matters, the vibronic exciton model developed in our previous work, and employed in the present article, emerges as a powerful framework to study the fission of a photoexcited singlet state into pairs of triplet excitons. Application of this model to a broader range of fission materials will allow for its further validation, and could highlight new physical factors that impact singlet fission beyond that in crystalline pentacene. The detailed insights into singlet fission so obtained promise to be a solid basis for a comprehensive understanding of this process.}

\section{Acknowledgements}

The authors thank Benedikt Kloss for helpful discussions. R.T.~acknowledges The Netherlands Organisation for Scientific Research NWO for support through a Rubicon grant. D.R.R.~acknowledges funding from NSF grant no.~CHE--1464802.

%


\end{document}